\documentclass[aps,preprint,superscriptaddress,]{revtex4}
\usepackage{graphicx,amsmath,amssymb,bm,placeins}
\usepackage{hyperref}

\begin{document}
\title{Topological origin of subgap conductance in insulating bilayer graphene}
\author{Jian Li}
\affiliation{D\'{e}partement de Physique Th\'{e}orique,
Universit\'{e} de Gen\`{e}ve, CH-1211 Gen\`{e}ve 4, Switzerland}
\author{Ivar Martin}
\affiliation{Theoretical Division, Los Alamos National Laboratory,
Los Alamos, New Mexico 87545, USA}
\author{Markus B\"{u}ttiker}
\affiliation{D\'{e}partement de Physique Th\'{e}orique,
Universit\'{e} de Gen\`{e}ve, CH-1211 Gen\`{e}ve 4, Switzerland}
\author{Alberto F. Morpurgo}
\affiliation{DPMC and GAP, Universit\'{e} de Gen\`{e}ve, CH-1211
Gen\`{e}ve 4, Switzerland}
\date{March 26, 2010}
\maketitle

\textbf{The edges of graphene-based systems possess unusual
electronic properties, originating from the non-trivial
topological structure associated to the pseudo-spinorial character
of the electron wave-functions. These properties, which have no
analogue for electrons described by the Schr\"odinger equation in
conventional systems, have led to the prediction of many striking
phenomena, such as gate-tunable ferromagnetism and
valley-selective transport \cite{son_half-metallic_2006,
rycerz_valley_2007, guinea_energy_2010}. In most cases, however,
the predicted phenomena are not expected to survive the influence
of the strong structural and chemical disorder that unavoidably
affects the edges of real graphene devices. Here, we present a
theoretical investigation of the intrinsic low-energy states at
the edges of electrostatically gapped bilayer graphene (BLG), and
find that the contribution of edge modes to the conductance of
realistic devices remains sizable even for highly imperfect edges.
This edge conductance dominates over the bulk contribution if the
electrostatically induced gap is sufficiently large, and accounts
for seemingly conflicting observations made in recent transport
and optical spectroscopy experiments
\cite{oostinga_gate-induced_2008, zhang_direct_2009,
mak_observation_2009}. Our results illustrate the robustness of
phenomena whose origin is rooted in the topology of the electronic
band-structure, even in the absence of specific protection
mechanisms.}

The electronic structure of graphene single- and bi-layers
possesses nontrivial topological properties, which originate from
the spinorial nature of the electronic wave-functions, and that
manifest themselves in many ways. For instance, when transported
along a closed trajectory, the wave-function of an electron in
single- or bi-layer graphene acquires a non-vanishing Berry phase,
responsible for the specific quantization sequence of the Hall
conductance that is observed experimentally
\cite{novoselov_two-dimensional_2005, zhang_experimental_2005,
novoselov_unconventional_2006}. Another manifestation of the
topological properties is the presence of zero-energy states
living at the material edges \cite{nakada_edge_1996,
castro_localized_2008, yao_edge_2009}. The existence of these
states is allowed by the richer structure of edge boundary
conditions (as compared to the conventional scalar wave-functions)
that is associated to the spinorial nature of the electronic
wave-functions (see, e.g., \cite{brey_electronic_2006}).

Many fascinating phenomena originating from the properties of edge
states in graphene layers have been predicted theoretically, such
as electric-field tunable magnetism \cite{son_half-metallic_2006}
and valley-dependent transport \cite{rycerz_valley_2007,
guinea_energy_2010}. In most cases, however, the possibility to
observe these phenomena is dubious, because theory often relies on
a perfectly ordered atomic structure of the graphene edges,
whereas in real materials strong disorder is unavoidably present.
To examine if exotic aspects of the edge physics can survive
disorder, here we analyze the transport properties of low-energy
edge states in the case of gapped BLG
\cite{mccann_landau-level_2006}. Contrary to the naive
expectation, we show that transport mediated by the low-energy
edge states exhibits universal features, and persists even in the
presence of very strong disorder.

To illustrate the topological origin of edge states in gapped BLG
(Fig. 1a) we describe the low-energy  electronic properties of
this material in terms of an effective (dimensionless)
Hamiltonian, by now well-established both theoretically and
experimentally \cite{mccann_landau-level_2006}:
\begin{eqnarray} \label{eq:ham}
&H_\tau = -\bm{g}_\tau(k_x,k_y)\cdot\bm{\sigma},& \\
&\bm{g}_\tau(k_x,k_y) = (k_x^2-k_y^2,\, 2k_xk_y\tau,\,
\Delta)),&\nonumber
\end{eqnarray}
where $\bm{\sigma}$ stands for the vector of Pauli matrices, $\tau
= \pm 1$ indexes the two valleys in the band-structure
(corresponding to the K and K' points in the Brillouin zone),
${\bf k} = (k_x,k_y)$ is the momentum relative to the K/K' point,
and $2\Delta$ defines the size of the gap. This Hamiltonian --
which can be derived as a long-wavelength low-energy limit of a
tight-binding description -- acts on spinors, whose components
correspond approximately to the amplitude of the wave-function on
atoms A1 and B2 of the bilayer unit cell (see Supporting
Information for details). The non-trivial topological properties
of BLG are described by the topological charge
\cite{volovik_universe_2003, martin_topological_2008}
\begin{equation} \label{eq:topo}
c_\tau = \frac{1}{4\pi}\int dk_x \int dk_y \:\hat{\bm{g}}_\tau \cdot \left(
\frac{\partial \hat{\bm{g}}_\tau}{\partial k_x} \times \frac{\partial
\hat{\bm{g}}_\tau}{\partial k_y} \right),
\end{equation}
which characterizes the mapping from a the region of the Brillouin
zone that contains the valley point K/K' to the unit vector
$\bm{g}_\tau$ defining the single-valley Hamiltonian. The
topological charge $c_\tau$ equals 1 in one of the valleys and
$-1$ in the other. From the bulk--edge correspondence
\cite{qi_topological_2008, hasan_topological_2010}, it can then be
expected that zero-energy  states, propagating in opposite
directions in the two valleys, appear at the material edge as a
consequence of the discontinuity of the topological charge between
BLG and vacuum. Indeed, ``valley-helical" states appear at
interfaces separating regions of BLG where the gap has opposite
polarity \cite{martin_topological_2008}, and have been shown to
exist at zig-zag edges by Castro {\it et al.}
\cite{castro_localized_2008} (Note that the latter authors based
their analysis on the full tight-binding description of graphene,
but the same states can be found in terms of the continuum
low-energy Hamiltonian (1); see Fig. 1b and Supporting
Information). In  a device consisting of a strip of BLG with
zig-zag edges, these zero-energy states propagate ballistically
and are responsible for a finite conductance equal to $4e^2/h$
(corresponding to two spin degenerate states at each one of the
device edges) when the Fermi level is located inside the gap.

These properties of  BLG parallel those of spin-orbit induced
topological insulators \cite{kane_quantum_2005, kane_z2_2005,
bernevig_quantum_2006, hasan_topological_2010}, in which ballistic
helical edge states have been observed experimentally
\cite{konig_quantum_2007, roth_nonlocal_2009}, but with the valley
quantum number playing the role of the electron spin. A complete
analogy between BLG and spin-orbit induced topological insulators,
however, is prevented by a number of important differences. Most
crucially, edge states in spin-orbit induced topological
insulators cannot back-scatter (and hence do not localize) as long
as time reversal symmetry is present -- a phenomenon referred to
as topological protection \cite{kane_quantum_2005, kane_z2_2005,
bernevig_quantum_2006}. In graphene bilayers, on the other hand,
any mechanism coupling states in different valleys can affect the
edges states, since intervalley scattering is equivalent (in our
analogy) to spin-flip processes for spin-orbit induced topological
insulators. For instance, at an ideal armchair edge where the
valleys are coupled by the boundary conditions
\cite{brey_electronic_2006}, the zero-energy edge states are
absent. For the same reason, it might be expected that any source
of short range scattering due to the edge non-ideality will have
the same effect, making the behavior found for ideal zig-zag edges
irrelevant for the description of transport in real devices.

To analyze the properties of the low-energy states present at disordered
edges we go back to the tight-binding description of gapped BLG,
and perform numerical calculations of conductance for strips of
gapped BLG. Short-range disorder is introduced in two different
ways, which model different physical phenomena occurring at real
graphene edges. Structural edge disorder is generated by removing
randomly tight-binding sites in the two layers
\cite{mucciolo_conductance_2009}, up to a depth $d$ from the
starting edge (see examples in Fig. 1c and d). The effect of
generic chemical species binding to the outermost carbon atoms is
modeled by adding to the corresponding sites in the tight-binding
Hamiltonian an energy that is distributed randomly in the range
$\pm 1$ eV (i.e., the characteristic order of magnitude associated
to chemical bonds). We perform calculations starting from
different ``ideal" edges, including zig-zag, armchair, and other
edges generated by cutting the BLG along arbitrary
crystallographic directions, for many different microscopic
realization of disorder, from which we extract the
ensemble-averaged conductance. In the calculations, we use the
known values for the in- and out-of-plane hopping parameters of
BLG, and gap values in the range that is accessible to experiments
(see Methods Section).

The ensemble-averaged conductance calculated as a function of
sample length $L$ for different values of $d$ is shown in Fig.
\ref{fig:decay}a and b, corresponding to the cases of starting
ideal zig-zag and armchair edges, respectively. It is apparent
that in all cases the average conductance decays exponentially
with $L$. The calculated conductance is
independent of the width $W$ of the graphene strip, indicating
that transport is indeed occurring along the edges (see Fig.
\ref{fig:decay}d). Only for very short strips ($L<5$ nm), a
sizable ``bulk" contribution to the conductance (scaling linearly
with the width of the strip) is also significant, due to direct
tunneling under the gap. This contribution decays very rapidly
with increasing the device length, and can be neglected in the
range of values of $L$ that we are considering. Note also that the
conductance calculated for large values of $L$ extrapolates in all
cases to $4e^2/h$  for $L\rightarrow 0$, as expected for edge
transport (the factor 4 accounts for two edges and two spin directions).
Therefore, the observed exponential decay of $G$ as a function of
$L$ enables us to directly extract the localization length
$\lambda_{loc}$ of the edge states.

When starting from an ideal zig-zag edge, structural disorder
causes localization of the pre-existing zero-energy states, and
the observed decrease of $\lambda_{loc}$  with increasing $d$ is
expected (see Fig. \ref{fig:llvgs}; the case where only structural
disorder is considered corresponds to the points connected by the
continuous line). Unexpectedly, however, we find that for strong
disorder the localization length tends to saturate to a rather
large value, that depends only on the magnitude of the gap (see
Fig. \ref{fig:llvgs} inset). The same behavior is observed when
starting from other periodic edges, obtained by cutting the BLG
lattice along different crystallographic directions. The situation
is different when starting from an ideal armchair edge, as in that
case we observe that both the conductance and localization length
increase upon increasing the disorder strength. The increase
occurs because for an ideal armchair edge no states exist in the
gap, and disorder introduces low-energy states at the edges that
mediate transport. Remarkably, also in this case, at large
disorder strength the localization length saturates to the value
found for disordered zig-zag edges. When chemical disorder is also
included (see Fig. \ref{fig:decay}c and points connected by the
broken line in Fig. \ref{fig:llvgs}), the localization length
saturates in all cases again to the same universal value, already
for a very small amount of structural disorder ($d=1$).
Physically, this implies that in the presence of sufficiently
strong disorder there is no relation between the structure of the
starting (ideal) edge and the low-energy electronic properties. In
other words, our calculations show that in the presence of
sufficient disorder, may it be structural or chemical, the
electronic properties of graphene edges are universal, and they
support low-energy electronic states responsible for sub-gap
conduction.

Finding that low-energy sub-gap edge states -- with a very long
localization length given the large strength of disorder
considered -- are present and characterized by universal
properties is remarkable. To highlight the importance of the band
structure topology in the case of gapped BLG with disordered
edges, we have performed numerical calculations on a gapped
half-filled square lattice (see Fig. \ref{fig:gsl}a), which,
similarly to BLG, is bipartite, i.e., it is formed by two
interpenetrating (square) sublattices. A gap in the electronic
spectrum can be opened by unbalancing the sublattice energies (see
Supporting Information; we used values for the tight-binding
hopping amplitude and for the gap comparable to those of
graphene). Despite superficial similarity, unlike in BLG, the
resulting band structure is topologically trivial. Accordingly,
intra-gap states are neither expected nor observed numerically at
the ideal edges. Indeed, we find that for the square lattice,
transport in the ideal case is due to weak direct tunneling under
the gap. Edge roughness does not introduce edge states inside the
gap, but only further suppresses conductance (see Fig.
\ref{fig:gsl}b and c). This comparison underscores the fundamental
importance of the topological origin of low-energy edge states in
gapped BLG. Even though these edge states are not protected
against short-range potential scattering, as in the case for
spin-orbit induced topological insulators \cite{kane_quantum_2005,
bernevig_quantum_2006}, their robustness manifests itself in a
very long and universal localization length.

Having established that the presence of low-energy edge states
with long localization length is a generic property of gapped BLG,
we discuss how these states manifest in transport experiments. In
this regard,  the ability to tune the gap in BLG electrostatically
up to values  of $\approx 300$ meV \cite{zhang_direct_2009,
mak_observation_2009} is particularly useful. For large gap
values, hopping though localized states is expected to be the
dominant low-temperature transport mechanism. Since the states
localized at the edge naturally exist near the middle of the gap,
they will give a dominant contribution to the measured
conductance, as long as the potential fluctuations due to the bulk
disorder are significantly smaller than the bulk gap. In real
devices this condition is satisfied when the gap is sufficiently
larger than the root mean square of potential fluctuations in the
bulk, which for samples on SiO$_2$ substrate has been measured to
be approximately 20 meV \cite{martin_observation_2008,
kuzmenko_determination_2009}. To compare with experiments, we
estimate the characteristic energy scale that determines the
temperature dependence of the conductance. To this end, we assume
that transport is dominated by variable range hopping along the
edges (one dimensional transport; the case of nearest neighbor
hopping would give a comparable result). The temperature
dependence of the conductance can then be written in the  form $G
\propto \exp[-(T^*/T)^{1/2}]$, with $T^*\approx V_g t a/(t_\perp
\lambda_{loc})$ (see Supporting Information; here we have denoted
the band-gap $2\Delta$ with $V_g$, to emphasize that in the
experiments this parameter can be controlled, and is proportional
to the voltage applied between the gates). Therefore, $T^*$ is the
characteristic energy scale that determines the low temperature
transport behavior: like the band gap, it increases with the
applied gate voltage but it is approximately one order of
magnitude smaller for the practically relevant values of $V_g$.

The dominating contribution of the edge states to the conductance
resolves an apparent experimental controversy, namely that the
energy scale commonly observed in the dc transport experiments
\cite{oostinga_gate-induced_2008} is significantly smaller than
the bulk band gap expected theoretically or measured directly by
optical spectroscopy \cite{zhang_direct_2009,
mak_observation_2009}. Indeed, in their study of the insulating
state of electrostatically gapped BLG Oostinga {\it et al}
\cite{oostinga_gate-induced_2008} extracted  a characteristic
transport energy scale of about 5-10 meV, whereas the largest
perpendicular electric field applied in that experiment
corresponded to a band-gap of approximately 100-120 meV (as it can
be inferred from the optical spectroscopy work performed later
\cite{zhang_direct_2009, mak_observation_2009}). Very recently,
similar experiments have also been successfully performed on
suspended, double-gated BLG (i.e., with the bulk of the bilayer
not affected by the direct contact with other materials)
\cite{allen_tunable_2010}. In that case the maximum gap that could
be opened electrostatically was estimated to be approximately 5
meV, while the resistance measured in dc transport exhibited an
insulating behavior with an activation energy of 0.3 meV. For both
experiments, the difference between expected bulk gap and
characteristic energy scale measured in transport is about one
order of magnitude, consistent with our analysis. The fact that,
experimentally, agreement is found irrespective of whether the
bilayer is in contact with a substrate or suspended indicates that
the observed sub-gap transport is not likely to be caused by the
bulk disorder. We therefore conclude that existing experiments do
indeed support the scenario presented here, namely that sub-gap
transport is dominated by conductance at the edges of the graphene
bilayer.

It is quite surprising that low-energy edge states in gapped BLG
survive -- and dominate transport -- even in the presence
of very strong disorder, even though the topological protection
mechanisms, which operate in the spin-orbit induced
topological insulators, are absent. This finding illustrates the
robustness of phenomena rooted in the underlying bulk
band-structure topology, whose broad importance for condensed
matter physics is starting to be fully appreciated only now. A
unique feature of BLG is that, by using gate electrodes to control
the magnitude and sign of the gap between valence and conductance
band \cite{martin_topological_2008}, this underlying topology can
be manipulated experimentally in nano-electronic devices.

\textbf{Methods} The rough edges of BLG samples in our simulations
are generated by random walks (forward only) near the sample
boundaries up to depth $d$, and 1000 random configurations are
averaged for each data point. The calculations are always done at
$E_f=0$ for the samples, with highly doped contacts. The
two-terminal conductance is calculated by using the recursive
Green's function method, and the local density of states is
obtained from the spectral function. The important parameters for
the BLG samples, unless otherwise specified, are: $t=2.8$eV,
$t_\perp=0.1t$, $V_g=0.03t$, $a=1.42{\AA}$, and $W=160$ (in units
of number of atomic layers; inter-edge scattering is negligible
with this width).

\textbf{Acknowledgements} This work has been supported by the
Swiss National Science Foundation (projects 200020-121807,
200021-121569) and by the Swiss Center of Excellence MaNEP. The
work of IM was carried out under the auspices of the National
Nuclear Security Administration of the U.S. Department of Energy
at Los Alamos National Laboratory under Contract No.
DE-AC52-06NA25396 and supported by the LANL/LDRD Program.

\pagebreak


\pagebreak \FloatBarrier

\begin{figure*}[htbp]
\begin{center}
  \includegraphics[width=0.7\textwidth]{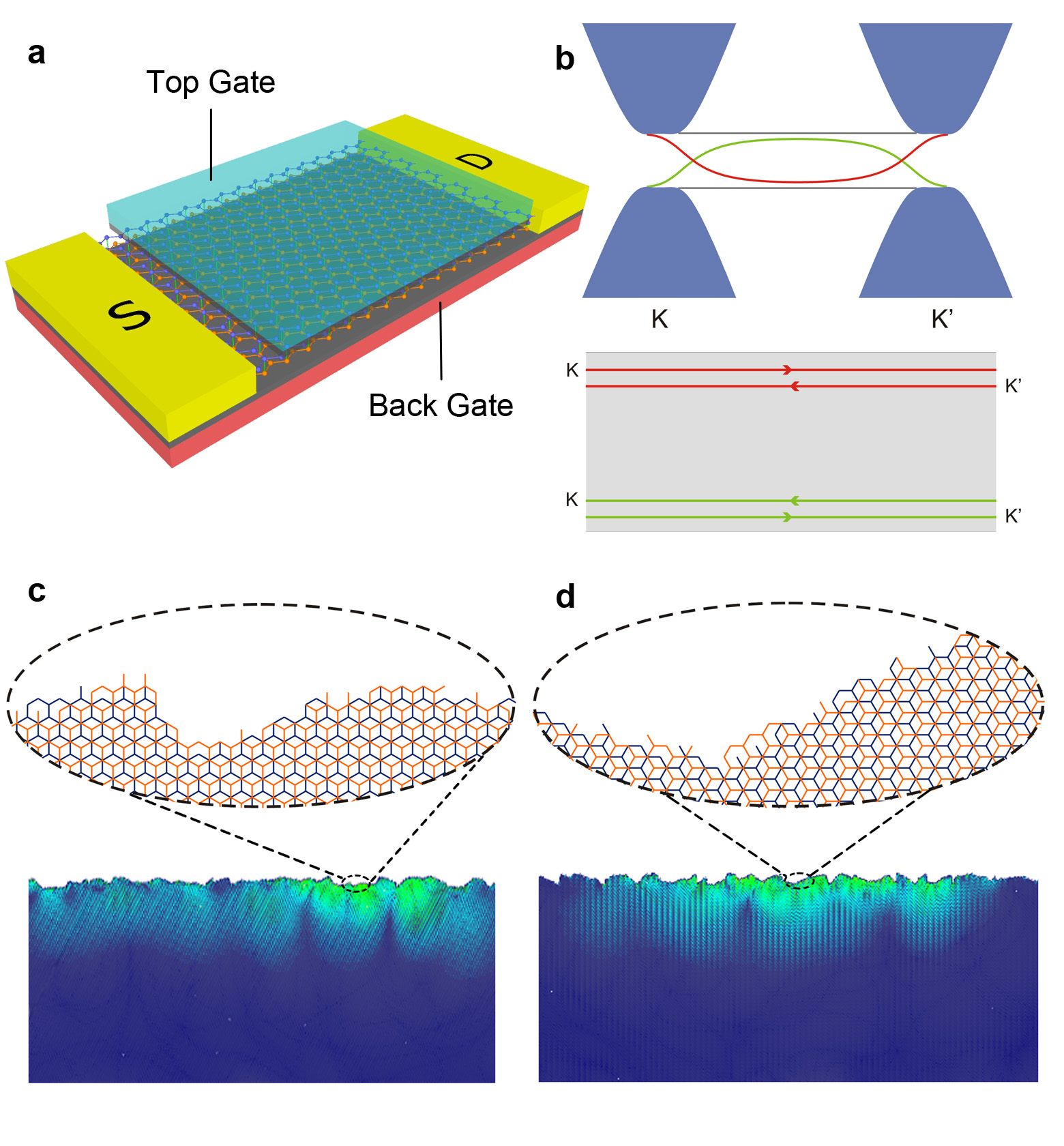}
\end{center}
\caption{\textbf{Edge states in gapped BLG.} Panel \textbf{a}
shows a schematic view of the device. By controlling the gate
voltages applied to the top and back gates separately, an energy
gap in the bulk of BLG can be opened and tuned, while maintaining
the Fermi energy in its center. Panel \textbf{b} shows the
dispersion of the sub-gap edge states in gapped BLG at a zig-zag
edge. At both edges, the states are helical with respect to the
valley degree of freedom, with states in opposite valleys
propagating in opposite directions. This conclusion holds for
periodic edges of rather general shape (i.e. not only for
zig-zag), but it does not hold for the ideal armchair edge, where
two valleys (K and K') are fully coupled and no sub-gap edge
states are present. Panels \textbf{c} and \textbf{d} show the
typical probability density near zero energy for strongly
disordered BLG zig-zag (\textbf{c}) or armchair (\textbf{d})
edges. In the presence of realistically strong disorder, the
existence of such edge states -- and their long localization
length -- is  a universal property of gapped BLG (the panels
correspond to a BLG that is approximately 100 nm long).}
\label{fig:dev_edge}
\end{figure*}

\pagestyle{empty} \linespread{1.3}

\begin{figure*}
\begin{center}
  \includegraphics[width=0.8\textwidth]{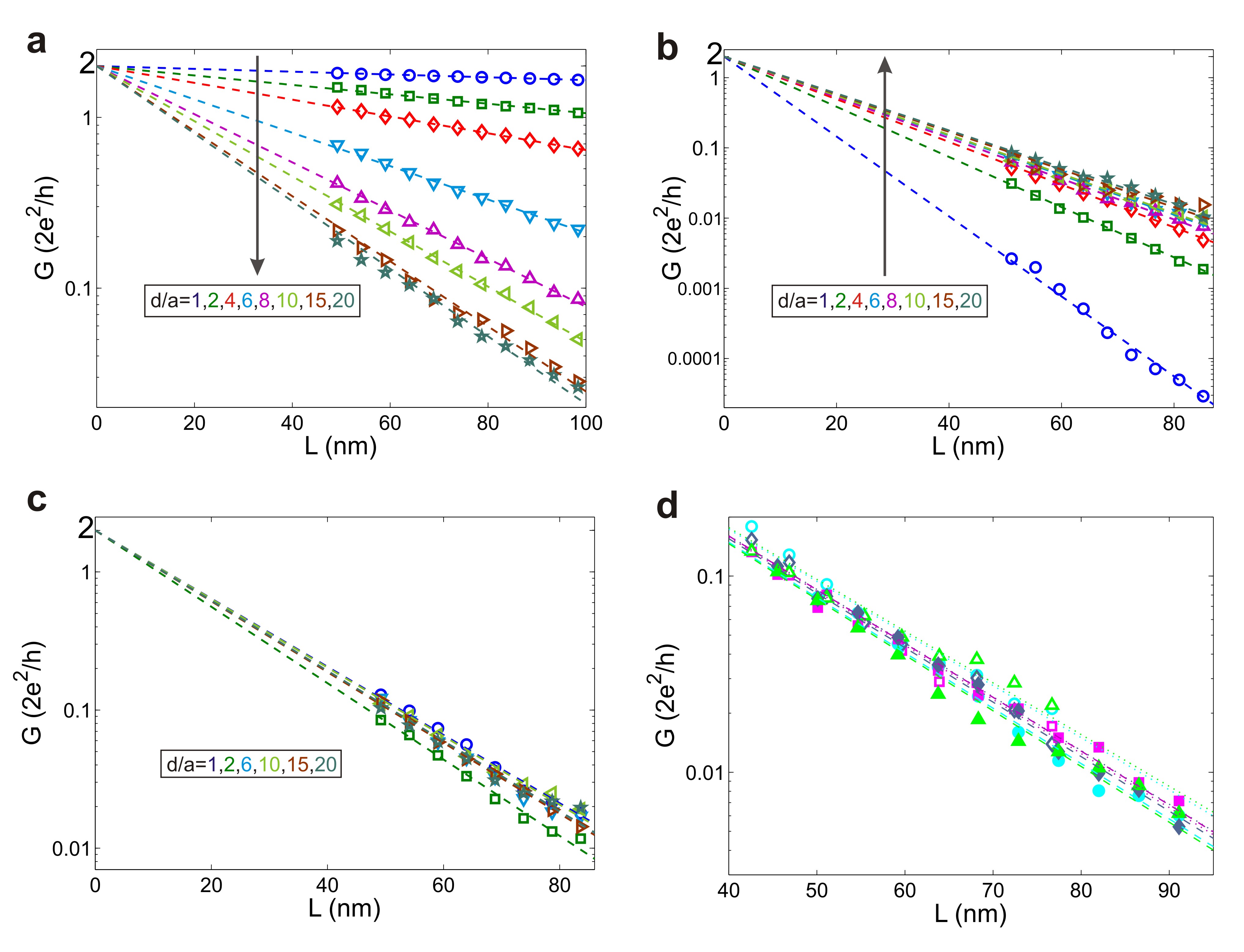}
\end{center}
\caption{\textbf{Subgap conductance in disordered BLG.} Panels
\textbf{a}-\textbf{d} show the decay of average conductance $G$
(at $E_f=0$) with increasing length $L$ of the BLG strips, for
different amounts of disorder. In panel \textbf{a} and \textbf{b},
the arrows indicate that the evolution of the conductance with
increasing structural edge roughness (quantified by the edge
roughness depth $d$) is opposite for samples starting from ideal
zig-zag (\textbf{a}) and armchair (\textbf{b}) edges. The
functional dependencies of the conductance as a function of $L$
calculated for different values of $d$ converge to the same
universal limit when disorder is sufficiently strong (i.e., the
$G(L)$ traces become universal at large  $d$). Panel \textbf{c}
shows that in the presence of ``chemical" disorder --consisting of
random on-site energies for the outermost tight-binding sites,
distributed between $\pm 1$ eV-- the $G(L)$ traces become rather
insensitive to the value of $d$ (i.e., chemical disorder or large
structural disorder have identical effects; the data shown in the
panel are obtained starting from an ideal zig-zag edge). Panel
\textbf{d} further shows the comparison of $G(L)$ traces
calculated in the presence of strong disorder (in the universal
limit) starting from armchair edges and for another
crystallographic edge obtained by cutting the BLG edge at a
$\theta\approx 10.9^\circ$ angle from the armchair direction
(calculations were performed with $d=20 a$). It is apparent that
the results are virtually identical, confirming once again that
for strong disorder different orientations become equivalent. The
data shown in panel \textbf{d} have been calculated for samples of
different width $W$ (in units of number of atomic layers: circles
correspond to $W=160$, squares to $W=200$, diamonds to $W=240$,
triangles to $W=280$): the lack of sensitivity to $W$ indicates
that sub-gap transport is dominated by edge states and not by the
BLG bulk. The fact that transport is occurring at the edges can
also be inferred from the data of panels \textbf{a-c}, that show
that the conductance extrapolates in all cases to $G\rightarrow 2$
(in units of $2e^2/h$) in the limit $L\rightarrow 0$
(corresponding to one spin degenerate edge channel at each sample
edge).} \label{fig:decay}
\end{figure*}

\linespread{1.6}

\begin{figure*}
\begin{center}
  \includegraphics[width=0.55\textwidth]{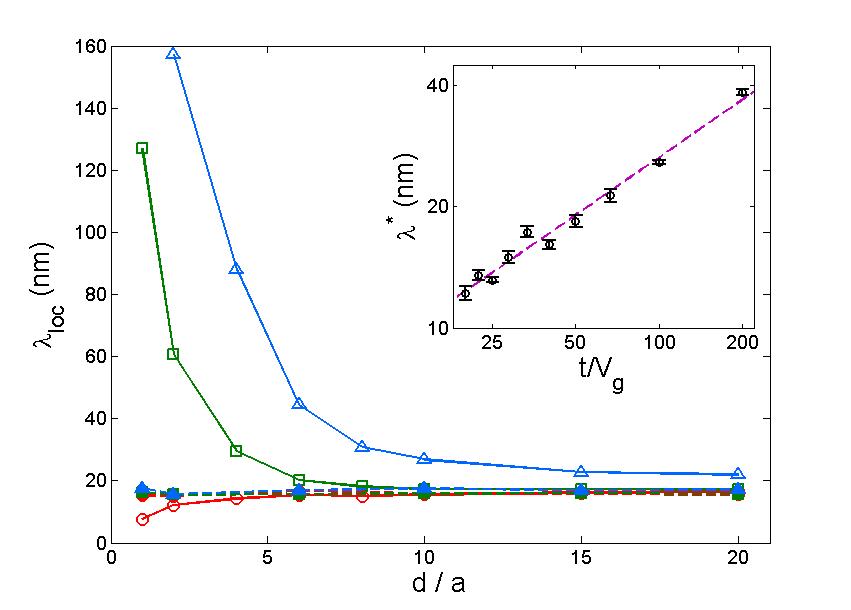}
\end{center}
\caption{\textbf{Universal behavior of localized edge states.} The
figure shows the dependence of the localization length
$\lambda_{loc}$ on the amount of  edge disorder, for the different
orientations of the ideal ``starting" edges (the data are
extracted from plots similar to those shown in
Fig.\ref{fig:decay}: circles correspond to armchair edges, squares
to a direction cut at a $\theta\approx 10.9^\circ$ angle from the
armchair direction, triangles to zig-zag edges). The cases in
which only structural disorder is considered -- quantified by the
roughness depth $d$ -- correspond to the data points connected by
solid lines. In those cases $\lambda_{loc}$ depends on both $d$
and $\theta$, until it saturates at large $d$. The cases in which
strong ``chemical" disorder is also included, modeled by random on
site energies on the outermost row of carbon atoms, correspond to
the data points connected by the broken line. With chemical
disorder, the ``universal" value of the localization length is
reached already for very small structural disorder (i.e. $d\simeq
1$). In all cases, sufficiently strong disorder leads to the same
value of $\lambda_{loc}$ regardless of any details. This
demonstrates that, for sufficiently strong disorder, the
localization length of the edge states is universal, and depends
only on the magnitude of the band-gap $V_g$. The universal value
of $\lambda_{loc}$ for different values of the gap $V_g$ is shown
in the inset. We find that  $\lambda_{loc}$ is  approximately
proportional to $1/\sqrt{V_g}$ (see the log-log plot in the inset;
the line is a fit yielding $\lambda_{loc} \propto 1/V_g^{\nu}$
with $\nu = 0.48 \pm 0.06$).} \label{fig:llvgs}
\end{figure*}

\begin{figure*}
\begin{center}
  \includegraphics[width=0.8\textwidth]{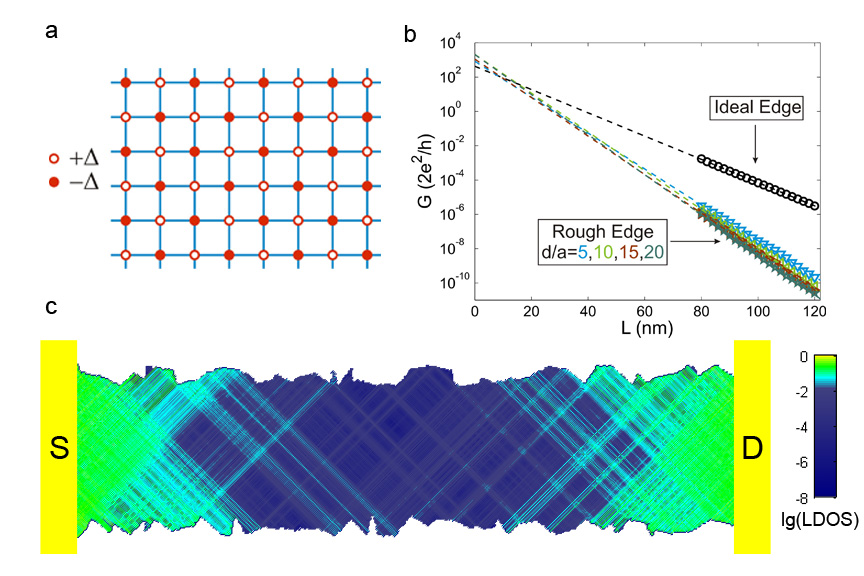}
\end{center}
\caption{\textbf{Absence of disorder-induced edge transport in a
topologically trivial system.} As a term of comparison for gapped
BLG, we analyze sub-gap transport in a square lattice, with a
staggered potential $\Delta$ that determines the size of the gap,
equal to $2\Delta$ (see panel  \textbf{a}; the lattice constant
$a$ is set to $a=1{\AA}$). Sub-gap transport is investigated by
setting the Fermi energy at the center of the gap. Panel
\textbf{b} shows the subgap conductance $G$ as a function of strip
length $L$, for the ideal edge, and for edges with different
amount of structural disorder. In this case edge roughness
significantly lowers the conductance as compared to the case of
ideal edges. In all cases, transport occurs through the bulk of
the system and is due to evanescent modes decaying from the source
and drain contacts (S and D in panel \textbf{c}), i.e. it is due
to tunneling through the bulk gap. This is why, contrary to gapped
BLG, the conductance extrapolates to large values when
$L\rightarrow 0$. Panel \textbf{c} shows the local density of
states in a rough-edge sample 60 nm long. The scar-like
trajectories in the bulk are caused by edge disorder, which affect
the evanescent waves that decay exponentially away from the
contacts. No evidence of edge states is found in this system.}
\label{fig:gsl}
\end{figure*}

\end{document}